\documentclass{iaus}
\usepackage{graphicx}

\title[] 
{Chemical composition of stellar populations in $\omega$ Centauri}

\author
 [Marino et al.]
{A. F. Marino$^{1,2}$, G. Piotto$^1$, R. Gratton$^3$, A. P. Milone$^1$, M. Zoccali$^2$, L. R. Bedin$^4$, S. Villanova$^5$ and A. Bellini$^{1,4}$}   

\affiliation{$^1$Dept. of Astronomy, University of Padova,
   Vicolo dell'osservatorio 3,
35122, Padova, Italy 
\\ email: {\tt  anna.marino@unipd.it} 
\\[\affilskip]
{ $^2$P. Univ. Cat\'olica de Chile, Dept. de
           Astronom\'ia y Astrof\'isica, Casilla 306, Santiago 22,
           Chile 
}

{$^3$INAF-Osservatorio Astronomico di Padova, Vicolo dell'Osservatorio 5, 35122 Padova, Italy 
}
 
{$^4$Space Telescope Science Institute, 3700 San Martin Drive,
Baltimore, MD 21218, USA 
}

{$^5$      Dept. de Astronomia, Universidad de Concepcion, Casilla
           160-C, Concepcion, Chile\\
}

}

\pubyear{2010}
\volume{268}  
\jname{Light elements in the Universe}
\editors{C. Charbonnel, M. Tosi, F. Primas \& C. Chiappini, eds.}
\begin{document}

\maketitle

\begin{abstract}
We derive abundances of Fe, Na, O, $\alpha$ and $s$-elements from 
GIRAFFE@VLT spectra
for more than 200 red giant stars in the Milky Way satellite $\omega$ Centauri.
Our preliminary results are that: ({\it i}) we confirm that $\omega$
Centauri exibiths large star-to-star metallicity variation ($\sim$ 1.4
dex); ({\it ii}) the metallicity distribution reveals the presence of
at least five stellar populations with different [{\rm Fe/H}]; 
({\it iii}) a distinct  Na-O
anticorrelation is clearly observed for the metal-poor and 
metal-intermediate stellar populations 
while apparently the anticorrelation disappears for the most
metal rich populations.
Interestingly the Na level grows with iron.

\keywords{Globular clusters, Stellar populations, Omega Centauri.}
\end{abstract}

\firstsection 
\section{Introduction}
Omega Centauri ($\omega$ Cen) is among the most studied and enigmatic
Milky  Way satellites.  It  has always been considered  a Globular
Cluster  (GC), but  a  number  of peculiarities,  like  the mass,  the
kinematics, and the complexity  of its numerous populations identified
by both spectroscopic and photometric investigations (Lee et al. 1999,
Pancino et al. 2000, Bedin et  al. 2004, Piotto et al. 2005, Villanova
et al. 2007),  suggest that it may be the remnant  of a larger stellar
system (Bekki \& Norris 2006).

Recently, it  has been shown  that some peculiar features  observed in
$\omega$ Cen are shared with other  GCs: A bimodality in s elements is
present  also in  NGC  1851 (Yong  et  al. 2008)  and  M22 (Marino  et
al. 2009),  and intrinsic  variations in [Fe/H]  were detected  in M22
(Marino et al.\ 2009), and  M54 (Sarajedini \& Layden 1995; Bellazzini
et al.\ 2008;  Carretta et al., submitted to  ApJL).  However $\omega$
Cen still remains the most exceptional GC in the Milky Way as concerns
its complex of numerous populations and large chemical variations.

Here  we   present  preliminary  results  of  our   project  aimed  to
characterize  the evolutionary connections  of the  sub-populations in
$\omega$ Cen, by studying their chemical content.

\section{Observations and data reduction}
We analysed FLAMES/GIRAFFE HR09 and  HR13 spectra for a sample of more
than 200 red giant stars. Abundances for iron  are  obtained  from  an
equivalent width  analysis by using the  Local Thermodynamical Equilibrium 
program
MOOG (C. Sneden, PhD thesis),
while the  other elements are  measured by comparing  observed spectra
with synthetic ones. More details on the abundance measurements can be
found in Marino et al.\ (2008, 2009).

\section{Results}
\begin{figure}[ht!]
\begin{center}
 \includegraphics[width=1.8in]{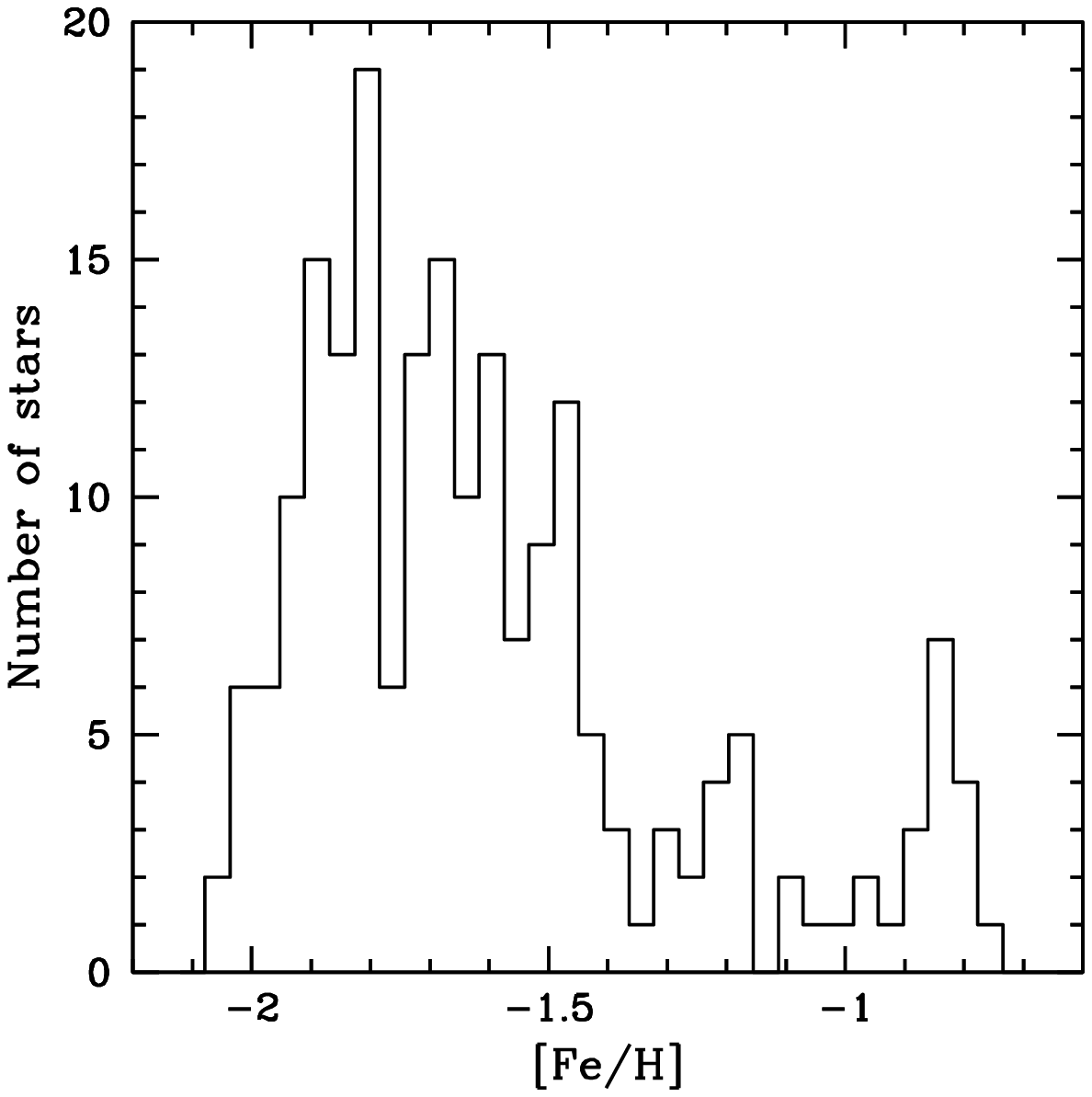}
 \includegraphics[width=3.4in]{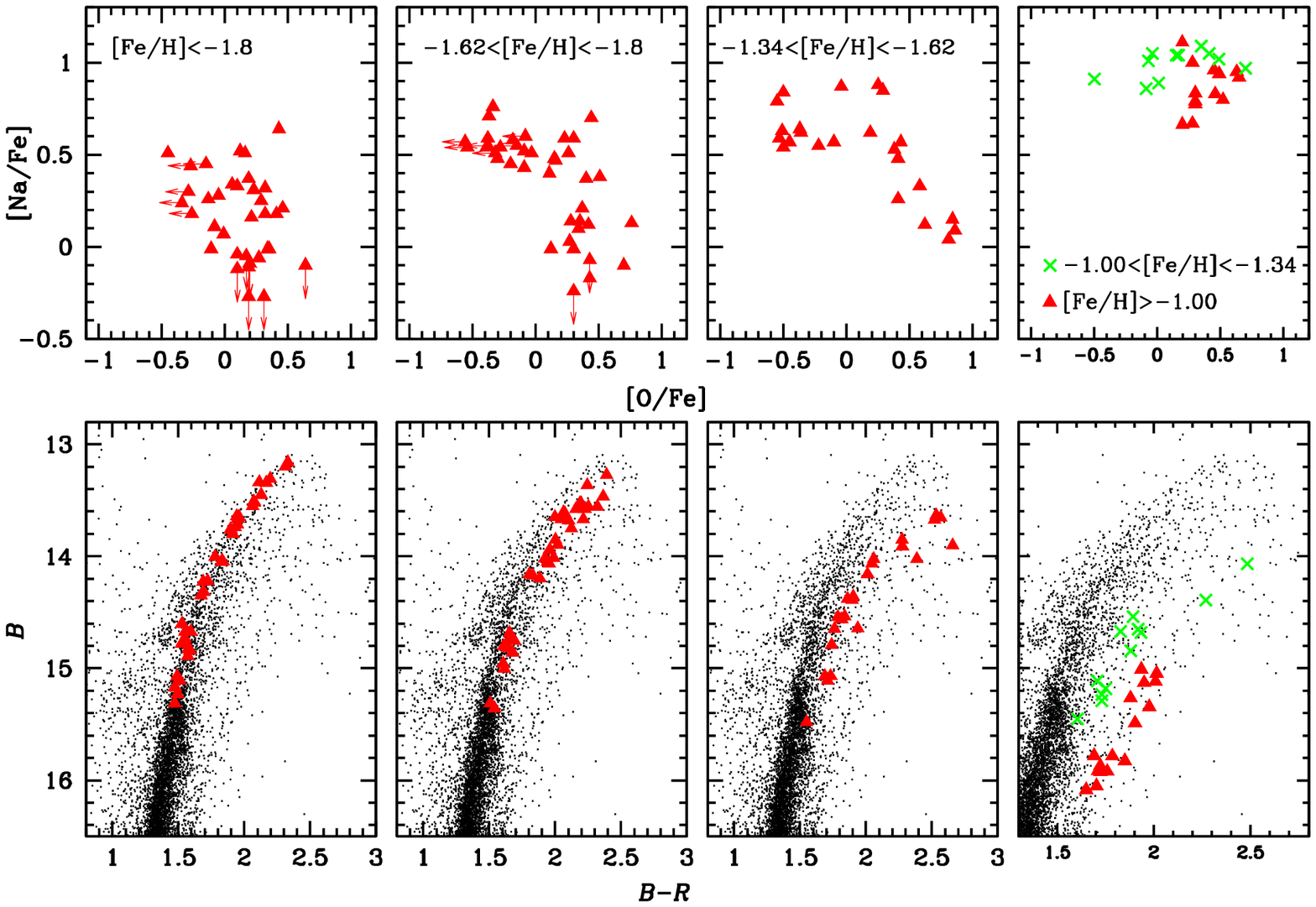}  
 \caption{\textit{Left panel}: Distribution of the iron content for our analysed sample. \textit{Right panels}: Na-O anticorrelation for the different groups of stars selected on the basis of the iron content and the position on the CMD represented in lower panels.}
   \label{NaO}
\end{center}
\end{figure}

We obtained  that the  [Fe/H] ranges from  $\sim -2.1$ to  $\sim -0.7$
dex, with  at least  five distinct peaks  in the iron  distribution as
shown in the left panel of Fig.\ref{NaO}.
In  the  right  panel of  Fig.\ref{NaO}  we  represent  the Na  and  O
abundances for  the five sub-populations in $\omega$  Cen, selected on
the basis  of different iron content  and their position  on the color
magnitude diagram (CMD). In lower panels  the position on  the $B$-$(B-R)$
CMD (Bellini et al. 2009)  for the different selected groups is shown,
with the  corresponding NaO anticorrelation  in each upper  panel.  We
note  that the  NaO anticorrelation  is  well defined  for the  stars
belonging to  the metal intermediate populations  (middle panels), and
some hints  of a probably less  extended one are present  for the more
metal poor  stars.  Apparently the anticorrelation  disappears for the
most metal rich populations.  Note  that the Na level grows with iron.

From the  analysis of the  s elements La and Ba  we derive that  the s
element abundance grows with increasing iron, and interstingly enough,
in each  group defined as  in Fig.\ref{NaO}, the s  element apparently
increase also with Na.

\end{document}